\documentclass{aa}
\usepackage{graphicx}
\def\gtrsim{\mathrel{\hbox{\rlap{\hbox{\lower4pt\hbox{$\sim$}}}\hbox{$>$}}}}
\def\ltrsim{\mathrel{\hbox{\rlap{\hbox{\lower4pt\hbox{$\sim$}}}\hbox{$<$}}}}

\def\etal {et al. }

\def\kmsM{{\rm\,km\,s^{-1}\,Mpc^{-1}}}
\begin{document}

\title{A critical appraisal of the SED fitting method to estimate 
photometric redshifts}

\author{M. Massarotti\inst{1,}\inst{2}
\and A. Iovino\inst{1} 
\and A. Buzzoni\inst{1,}\inst{3}}

\offprints{M. Massarotti}

\institute{Osservatorio Astronomico di Brera, Via Brera 28, 20121 Milano, Italy
\and
Universit\`a di Napoli, Dipartimento di Scienze Fisiche, Mostra d'Oltremare, 80125 Napoli, Italy
\and
Telescopio Nazionale Galileo, Roque de los Muchachos Astronomical Obs.,\\
P.O. Box 565, 38700 Santa Cruz de La Palma (TF), Spain\\
\email{massarot@brera.mi.astro.it; iovino@brera.mi.astro.it; buzzoni@tng.iac.es}}

\date{Received ; Accepted}

\abstract{
We discuss the stability of the photometric redshift estimate obtained
with the SED fitting method with respect to the choice of the galaxy
templates. Within the observational uncertainty and photometric
errors, we find satisfactory agreement among different sets of
theoretical and empirical templates using the Hubble Deep Field North
as a target galaxy sample. Our results suggest that, especially at
high redshift, the description of the physical processes of photon
absorption in the interstellar and intergalactic medium plays a
dominant role in the redshift estimate. The specific choice of the
template set, as long as this includes both normal and starburst
galaxies, is in comparison a minor issue.
}

\maketitle

\section{Introduction \label{s_i}}

Physical properties of high--redshift galaxies have been extensively
explored in recent years thanks to a number of deep spectroscopic
surveys carried out with the largest ground optical telescopes
(e.g. Lilly \etal1995; Cowie \etal1996; Ellis \etal1996). These
surveys have enabled to trace well galaxy evolution up to $z \sim
1.5$.

Beyond this redshift limit, however, ground--based spectroscopy is
forcedly confined to the brightest objects, a few magnitudes above the
limit achieved by deep photometry. For the time being, spectra of
galaxies fainter than $I \sim 24$ AB mag are still difficult to
obtain, and one must therefore rely on multicolor observations coupled
to photometric redshift techniques to estimate redshifts for the
majority of galaxies in this magnitude range.

Photometric redshift techniques have already been applied by many
authors to the ultra--deep images in the Hubble Deep Field North
(HDFN, Williams \etal1996), addressing in fine detail the study of
galaxy luminosity function, star formation rate (SFR), UV cosmological
density, and clustering properties (Madau \etal1996; Connolly
\etal1997; Sawicki \etal1997; Franceschini \etal1998; Pascarelle
\etal1998; Connolly \etal1998; Miralles \& Pell\'o 1998; Arnouts
\etal1999).

Two main approaches have been used in the literature to measure
photometric redshifts: the so--called training--set method
(e.g. Connolly \etal1997; Wang \etal1999), and the spectral energy
distribution (SED) fitting method (e.g. Sawicki \etal1997; Giallongo
\etal1998; Ben\'{\i}tez 2000).

In the training--set method the relation between colors (sometimes
also magnitudes) and redshift is calibrated using a spectroscopic
galaxy sample. The accuracy of this method depends on the number and
completeness of the reference templates, and obviously increases when 
acquiring new spectroscopic data. It becomes less reliable at high
redshifts because of incompleteness and increasing redshift uncertainty
of the most distant reference galaxies (see Ben\'{\i}tez 2000 and
Fern\'andez--Soto \etal2000 for a full discussion of the method
drawbacks).

In the SED fitting method, on the other hand, photometric redshift is
obtained by comparing observed galaxy fluxes, $f^\mathrm{obs}_i$ at
the $i$--th photometric band, with a library of reference fluxes,
$f^\mathrm{templ}_i(z, T)$, depending on redshift and a suitable set
of parameters $T$, that account for galaxy morphological type, age,
metallicity, dust reddening etc. In this way, for each galaxy one can
perform a $\chi^2$ confidence test obtaining the values of $z$ and $T$
that minimize flux residuals between observations and reference
templates.  In addition to the best--$z$ estimate, the method
identifies an ``optimum template'', supplying information on galaxy
age and morphology, although special caution must obviously be taken
in using this piece of information, because of possible numerical
degeneracies in the parameter space.

In this paper we will discuss in some detail the performance of the
SED fitting method as a function of the template set chosen to
reproduce galaxy colors. The primary aim is to assess the effects of
different template properties on the numerical stability and physical
consistency of the method.  In our opinion, this ``sanity check'' must
be preliminary to any direct comparison of photometric and
spectroscopic redshift determinations, as different template sets can
give statistically similar results when applied to the same
spectroscopic subsample, but produce a different redshift distribution
on the entire photometric catalog (Yee 1998).

The plan of the paper is as follows. We will first recall in Sec. 2
some key steps in the method. The theoretical and empirical template
libraries are introduced in Sec. 3, while Sec. 4 presents results
obtained by comparing the different galaxy template sets. The role of
physical processes in the interstellar and intergalactic medium is
discussed in Sec. 5, where we show their relevant impact on redshift
estimate.  Sec. 6 finally summarizes our conclusions. 

\section{The SED fitting algorithm}

Before proceeding with our discussion, it might be useful to briefly
remind how the SED fitting algorithm works.

As a first step one has to convert galaxy observed magnitudes for each
$i$--th photometric band into incoming apparent flux,
$f^\mathrm{obs}_i(\lambda)$.  This is equivalent to reconstruct the
SED of target galaxies at very low spectral resolution by sampling
their luminosity at the effective wavelength of the photometric bands
available.

For each object, the sampled flux has then to be compared with the
reference spectral libraries of template galaxies,
$f^\mathrm{templ}_i(z, T)$, computing a $\chi^2$ merit function of the
fitting residuals such as

\begin{equation}
\label{form_2}
\chi^2=\sum_{i=1}^{N}{{[f^\mathrm{obs}_i-s\ f^\mathrm{temp}_i(z, T)]^2}\over{\sigma_i^2}}\,.
\end{equation}

\noindent
In the equation, $N$ is the number of photometric bands, and
$\sigma_i$ is the observational uncertainty in the $i$--th band. The
scale factor $s$ is chosen in such a way as to minimize $\chi^2$ for
each template:

\begin{equation}
\label{form_3}
s = {{\sum_{i=1}^N f^\mathrm{obs}_i\ f^\mathrm{temp}_i(z,T)/\sigma_i^2}\over
     {\sum_{i=1}^N f^\mathrm{temp}_i(z,T)^2         /\sigma_i^2}}\,.
\end{equation}

\noindent

We will search the galaxy redshift in the range $z \in [0.0,
5.0]$ with a step of 0.02.

As in the current literature, a further constraint has been added to
the fit imposing that the age of the best template cannot be older
than the Hubble time at the selected redshift (in our calculations we
assumed a $H_0=50\kmsM$, $q_0=0.0$ cosmology with
$\Lambda = 0$). It can be verified, however, that redshift estimates
do not depend critically on the details of the cosmological model, and
this constraint could even be dropped with no serious impact on the
final output.

\section{Theoretical vs. empirical galaxy templates}
 
A key issue in the SED fitting method is the selection of the
reference flux library. In particular, one has to decide whether it is
more appropriate to use empirical or synthetic galaxy templates.

The major advantage of the first choice is that the observed SEDs for
a suitable set of local galaxies, spanning the whole range of Hubble
morphological types, gives {\it by definition} a physically consistent
picture of the real galaxies, at least the nearby ones.  On the other
hand, there is no obvious argument supporting a straightforward
extension of the {\it current} galaxy properties to
high--redshift objects. At earlier cosmological epochs, evolution
could play a substantial role in changing both morphological and
spectrophotometric properties of distant galaxies (see Buzzoni 1998
for a critical discussion on the subject).

In this respect, synthesis models should be preferred, as they do try
to take into account evolution. Of course, some differences exist
among current theoretical codes, especially in the SED predictions for
the UV range (Charlot \etal1996), and these differences should be
carefully examined when trying to match high--redshift observations to
derive $z$ from multicolor photometry. Optical (or even infrared)
observations of high-redshift objects sample UV restframe emission.

\begin{figure}
\resizebox{\hsize}{!}{\includegraphics{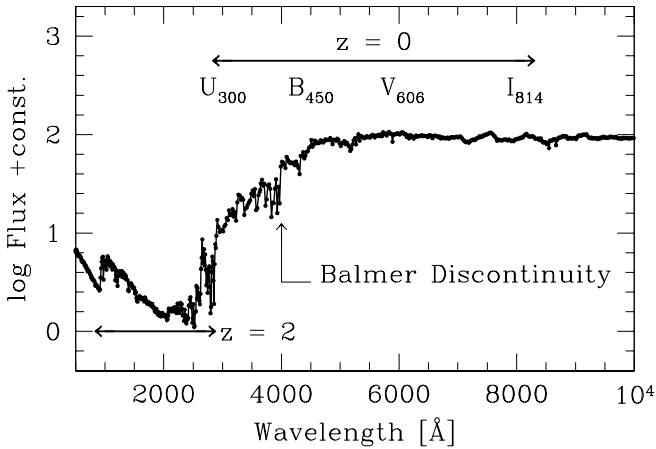}}
\caption{Restframe portion of galaxy SED explored by the WFPC2
photometric system at $z = 0$ and $z = 2$. As an illustrative example
a template model for local elliptical galaxies is shown.}
\label{u8}
\end{figure}

We should also recall that at high $z$ the restframe
wavelength range explored by a given photometric system becomes narrower. 
For example, at $z \sim 1$ the whole HST WFPC2 photometric
system (i.e. $U_{300}, B_{450}, V_{606}$ plus the $I_{814}$ band) is
just probing the wavelength range shortward of the galaxy 4000 \AA\
Balmer break, while for $z \sim 2$ the spectral window shrinks even
more to a scarce $\pm 800$ \AA\ interval around 1800 \AA\ in the
galaxy restframe (cf. Fig.~\ref{u8}).  Note in addition that, at least
for early--type galaxies, SED drops to nominal values shortward of the
Balmer discontinuity, so that $\kappa$ corrections become increasingly
positive with $z$. As we can see in Fig.~\ref{u8}, the effect can
reach up to two orders of magnitude for elliptical galaxies, and this
can result in a strong bias against their detection at high redshifts.

Another problem to be addressed when comparing photometric
observations of high--redshift galaxies with reference templates is
the role of intervening interstellar (ISM) and intergalactic medium
(IGM), which strongly modulates the observed galaxy SED.  Interstellar
dust attenuates the galaxy luminosity by an amount that greatly
depends on the specific galaxy environment (Gordon \etal 1997; Bruzual
\etal 1988; Kuchinski \etal 1998). The IGM, on the other hand,
drastically absorbs galaxy emission shortward of the Ly--$\alpha$
limit (the well known Ly--$\alpha$ forest effect, e.g. Madau
1995).  Unfortunately, present knowledge of these physical processes
does not yet permit any firm and univocal correction of the data,
leaving room for a variety of operational approaches (Sawicki
\etal1997; Giallongo \etal1998; Miralles \& Pell\'o 1998).

In our study we will include three sets of theoretical galaxy models,
namely those provided by the code of Bruzual \& Charlot (1993,
hereafter BC), Buzzoni (1989, 1995, 2000; BUZ) and Fioc and
Rocca--Volmerange (1997; FRV), as well as the empirical template set
of Coleman \etal (1980; CWW) extensively used by previous authors
(Sawicki \etal1997; Fern\'andez--Soto \etal1999, hereafter FSLY;
Ben\'{\i}tez 2000; Thompson \etal2000).

\subsection{Template libraries} 

For each galaxy we are mainly interested in a measure of its redshift,
but an age estimate and a morphological classification parameter are
also a result of the SED fitting technique. In order to ensure an 
homogeneous sampling of the galaxy distinctive parameters, a selected
grid of reference models for each of the three synthetic libraries has
therefore been considered.

According to the leading physical processes in the galaxy stellar
populations we can identify two main evolutionary regimes that induce
relevant differences in galaxy luminosity:

{\it i) ``Early'' galaxy evolution} - In the first Gyr of life, most
of the galaxy luminosity is provided by short-living stars of high
mass ($M \geq 2 M_\odot$, Buzzoni 2000). These are strong UV emitters
and dominate galaxy luminosity at short wavelength. Given their short
lifetime, these stars track actual star formation activity in the host
galaxy through a measure of its integrated UV flux.  Synthesis models
all converge toward a unified starburst scenario for $t < 1$ Gyr,
despite any later morphological evolution that could then
differentiate among spirals, ellipticals and irregulars (Buzzoni 1998;
Leitherer \etal1999).

{\it ii) ``Late'' galaxy evolution} - It comprises all the objects
older than 1 Gyr, displaying morphological features accounted for by
the standard Hubble classification.  Disregarding any specific detail
of star formation and morphology, the stellar bulk in old galaxies is
dominated by stars with $M \leq 2 M_\odot$ experiencing a full
red--giant--branch (RGB) phase tipping with a degenerate--core Helium
flash (Renzini \& Buzzoni 1986).

For each of our synthesis codes, we first set up an ``old--galaxy''
template library, tuning up reference models for present--day galaxies
along the whole Hubble sequence, and consistently tracking
back--in--time evolution up to $t = 1$ Gyr.  For ages less than 1 Gyr
all our synthesis codes have been complemented by the starburst models
at $t = 50$, 100, 500, and 800 Myr from Leitherer \etal(1999), better
suitable to describe high--mass stellar evolution.

Note that having used Leitherer \etal(1999) starburst models to
describe young galaxies does not influence our results, as all the
codes do actually converge, for early ages, to the redshift results
provided by the Leitherer \etal(1999) models.

An important issue to optimize computing resources and prevent
redundant oversampling in the parameter space is the age step in the
grid of template models.

\begin{table*}
\caption{Galaxy templates from the reference synthesis codes and from
Coleman et al. (1980)} \scriptsize
\begin{tabular} {lcccccll}
\hline
\noalign{\smallskip}
Template & Age & $U-V$ & $B-V$ & $V-K$ & SFR & IMF$^{(*)}$ & \multicolumn {1}{c}{Remarks} \\
         & ($z = 0$) &      &      &      &     &     &         \\
\noalign{\smallskip}
\hline
\noalign{\smallskip}
\multicolumn{4}{l} {\bf Bruzual \& Charlot (1993):} & & & & \\
         &     &      &      &      &     &     &         \\
E/S0  &  12 Gyr & 1.58 & 1.00 & 3.22 & $\propto$ exp(-t/$\tau$), $\tau = 1$ Gyr & Sca & $Z = Z_\odot$ \\
Sab   &  12 Gyr & 1.09 & 0.81 & 2.95 & $\propto$ exp(-t/$\tau$), $\tau = 4$ Gyr & Sca & $Z = Z_\odot$ \\
Scd   &  12 Gyr & 0.78 & 0.66 & 2.74 & $\propto$ exp(-t/$\tau$), $\tau = 15$ Gyr & Sca & $Z = Z_\odot$ \\
Im    &  1.4 Gyr~ & 0.13 & 0.28 & 1.89 & const. & Sca & $Z = Z_\odot$ \\
         &     &      &      &      &     &     &  \\
\hline
         &     &      &      &      &     &     &         \\
\multicolumn{4}{l} {\bf Buzzoni (2000):} & & & & \\
         &     &      &      &      &     &     &         \\
E    &  15 Gyr & 1.48 & 0.94 & 3.27 & single burst & Sal & two--zone model,\\
         &     &      &      &      &     &     &  $L_\mathrm{bol}$ (halo:bulge) = (0.15:0.85) \\
         &     &      &      &      &     &     & $\overline{Z}\mathrm{(halo+bulge)} = Z_\odot$ \\
Sb   &  15 Gyr & 0.58 & 0.63 & 2.79 & $\propto t^{-0.5}$ (disk) & Sal & three--zone model,\\
         &     &      &      &      &     &     & $L_\mathrm{bol}$ (halo:bulge:disk) = (0.05:0.30:0.65) \\
         &     &      &      &      &     &     & $\overline{Z}\mathrm{(halo+bulge)} = Z_\odot$, $Z\mathrm{(disk)} = Z_\odot/3$ \\
Sc   &  15 Gyr & 0.36 & 0.53 & 2.62 & $\propto t^{-0.1}$ (disk) & Sal & three--zone model,\\
         &     &      &      &      &     &     & $L_\mathrm{bol}$ (halo:bulge:disk) = (0.03:0.15:0.82) \\
         &     &      &      &      &     &     & $\overline{Z}\mathrm{(halo+bulge)} = Z_\odot$, $Z\mathrm{(disk)} = Z_\odot/3$ \\
Im   &  15 Gyr & 0.16 & 0.43 & 2.39 & $\propto t^{+0.8}$ & Sal & $Z = Z_\odot/3$ \\
         &     &      &      &      &     &     &         \\
\hline
         &     &      &      &      &     &     &         \\
\multicolumn{4}{l} {\bf Fioc \& Rocca Volmerange (1997):} & & & & \\
         &     &      &      &      &     &     &         \\
E    &  15 Gyr & 1.40 & 0.91 & 3.19 & Schmidt law & R\&B & Schmidt parameters ($n, \nu^{-1}$) = (1, 0.2 Gyr), \\
         &     &      &      &      &     &     & $Z = Z_\odot$ \\
Sb   &  12 Gyr & 0.92 & 0.73 & 3.14 & Schmidt law & R\&B & Schmidt parameters ($n, \nu^{-1}$) = (1, 2.8 Gyr),\\
         &     &      &      &      &     &     & $Z = Z_\odot$ \\
Sc   &  12 Gyr & 0.59 & 0.58 & 2.94 & Schmidt law & R\&B & Schmidt parameters ($n, \nu^{-1}$) = (1, 10.0 Gyr),\\
         &     &      &      &      &     &     & $Z = Z_\odot$ \\
Im   &  ~4 Gyr & 0.20 & 0.36 & 2.26 & Schmidt law & R\&B & Schmidt parameters ($n, \nu^{-1}$) = (1, 20.0 Gyr),\\
         &     &      &      &      &     &     & $Z = Z_\odot$ \\
         &     &      &      &      &     &     &         \\
\hline
         &     &      &      &      &     &     &         \\
\multicolumn{4}{l} {\bf Coleman et al. (1980):} & & & &  \\
         &     &      &      &      &     &     &         \\
E    &  - & 1.49 & 0.93 & - & - & - & M31 model \\
Sbc  &  - & 0.59 & 0.58 & - & - & - & - \\
Scd  &  - & 0.37 & 0.49 & - & - & - & - \\
Im   &  - & --0.06~~ & 0.32 & - & - & - & -  \\
         &     &      &      &      &     &     & \\
\hline

         &     &      &      &      &     &     &         \\
\multicolumn{4}{l} {\bf Leitherer et al. (1999):} & & & & \\
         &     &      &      &      &     &     &         \\
Starburst \#1 &  800 Myr & --0.27~~ & 0.13 & 1.62 & const. & Sal & $Z = Z_\odot$ \\
Starburst \#2 &  500 Myr & --0.37~~ & 0.09 & 1.62 & const. & Sal & $Z = Z_\odot$ \\
Starburst \#3 &  100 Myr & --0.66~~ & 0.02 & 1.70 & const. & Sal & $Z = Z_\odot$ \\
Starburst \#4 &  ~50 Myr & --0.76~~ & 0.00 & 1.72 & const. & Sal & $Z = Z_\odot$ \\
         &     &      &      &      &     &     &   \\
\hline
\end{tabular}
\begin{list}{}{}
\item $^{(*)}$ Sal = Salpeter (1955), Sca = Scalo (1986), R\&B = Rana \& Basu (1992).
\end{list}
\label{tbl_all}
\end{table*}

Based on the theory of simple stellar populations (cf. e.g. Buzzoni
1989) we know that luminosity evolution proceeds in this case like $L
\propto t^{-\beta}$, with $\beta$ varying in the range between unity
at UV wavelength and 0.7 in the infrared ($K$ band). Even accounting
for a smoother evolution, e.g. for a continuous SFR, a $\Delta\ t/t
\sim 0.3$ age sampling is sufficient to confidently appreciate any
0.1~mag change in apparent colors for high--$z$ spirals and
ellipticals.

The adopted BC templates consist of models with four different SFR
histories to match today's $E/S0$, $Sa/Sb$, $Sc/Sd$, and $Im$ Hubble
types. From the BUZ and FRV codes, we selected the original templates
for $E$, $Sb$, $Sc$, $Im$ morphological types. The BC models assume a
Scalo IMF, while the FRV models assume a Rana \& Basu (1992) mass
distribution, and the BUZ models adopt a standard Salpeter IMF.  For
all the models we sampled the different Hubble types at $t = 1.0, 1.4,
2.0, 4.0, 8.0, 10.0, 12.0$ and 15.0 Gyr.

Table~\ref{tbl_all} gives a complete summary of the relevant
distinctive properties of each theoretical library, reporting also the
expected colors for present--day galaxies of different Hubble type.

The original CWW empirical template set consisted of four observed
galaxy spectra, $E$, $Sbc$, $Scd$ and $Im$ respectively.  Sawicki
\etal(1997) and Ben\'{\i}tez (2000) had already noticed that adding
starburst templates (sensibly bluer than any $Im$ SED) to CWW greatly
alleviates the systematic discrepancies found in redshift distribution
of target galaxies beyond $z>1.5$. Similarly, the lack of starburst
templates could have biased FSLY results by inducing an artificial
peak in the galaxy distribution around $z \sim 1.7$. In this work,
following the strategy adopted by Sawicki \etal(1997), we extended the
CWW empirical library by adding two starburst models from Leitherer
\etal(1999) at $t = 50$ and $500$ Myr.  This enlarged dataset will
hereafter be referred to as the CWW--extended (CWWE) set.

\subsection{ISM and IGM corrections}

Fluxes provided by synthetic SED models have to be corrected for ISM
and IGM absorption effects.  Both are wavelength dependent, the latter
is also redshift dependent, and they can dramatically change the
galaxy spectrum in the UV restframe.

\begin{figure}
\resizebox{\hsize}{!}{\includegraphics{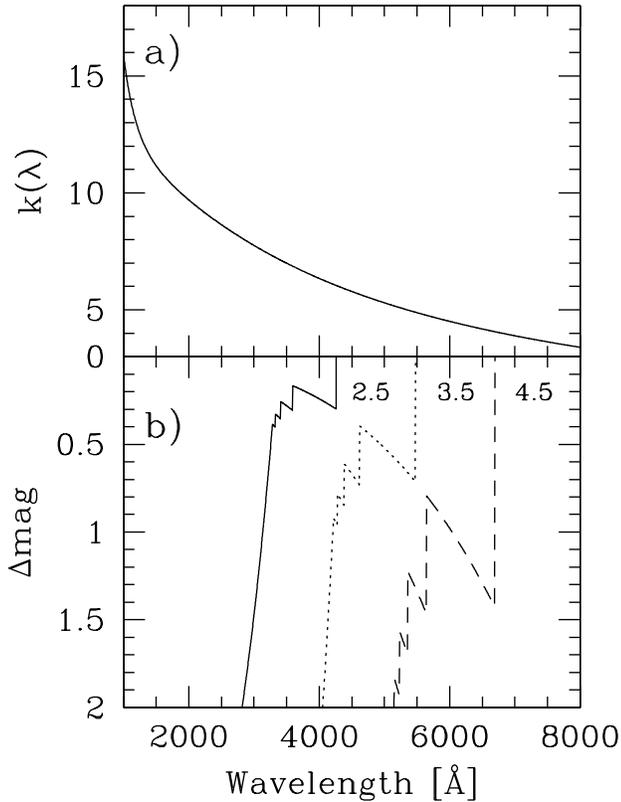}}
\caption{{\it Panel a:} the adopted dust attenuation law after
Calzetti (1999). {\it Panel b:} the IGM attenuation in a magnitude
scale at redshift 2.5, 3.5 and 4.5, as labelled, according to Madau
(1995).}
\label{fig-1}
\end{figure}

Interstellar dust attenuation depends on the color excess parameter
$E(B-V)$, scaling with dust column density, as well as on the
differential ISM opacity $k(\lambda)$, which is related to the
physical properties of dust grains.  The classical
relationship for intrinsic and emerging flux is:

\begin{equation}
\label{form_1}
F_\mathrm{obs}(\lambda)=F_0(\lambda)10^{-0.4E(B-V)k(\lambda)}\,.
\end{equation}
\noindent

In our work we adopted the Calzetti (1999) dust attenuation law
$k(\lambda)$ (see Fig.~\ref{fig-1}, panel {\it a}) including grain
absorption and scattering. The $E(B-V)$ parameter can take the
values 0.0, 0.05, 0.1, 0.2, 0.3. By definition, the empirical CWW
spectra already include internal reddening.

We describe the IGM absorption as a function of redshift following
Madau (1995) (Fig.~\ref{fig-1}, panel {\it b}).

\begin{figure}
\resizebox{\hsize}{!}{\includegraphics{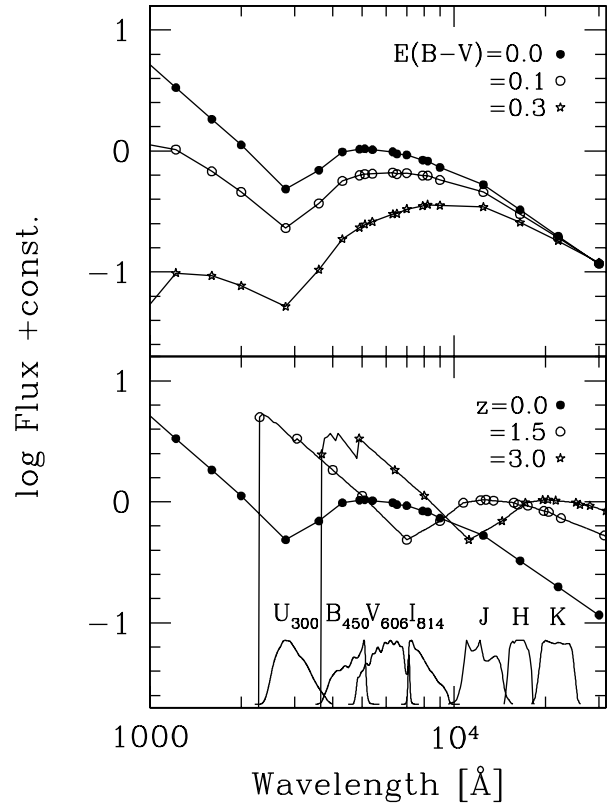}}
\caption{The effect of internal reddening and IGM on the Sb reference
template from Buzzoni (2000). Dust attenuation for $E(B-V)$ up to 0.3
mag, as labelled, is shown in the upper panel, while the expected
break induced by the Ly--$\alpha$ forest at $z = 1.5$ and $z = 3$ is
shown in the lower panel. For reference, the HST photometric
system and the Johnson $JHK$ bands are displayed at the bottom.}
\label{fig-100}
\end{figure}

To give an illustrative example of the combined action of dust and
IGM, in Fig.~\ref{fig-100} the BUZ Sb template has been reddened by
different amounts of the color excess $E(B-V)$ up to 0.3 magnitudes
(upper panel).  The effect of the IGM on the same spectrum is
displayed on the contrary in the lower panel for the two cases of
redshift $z = 1.5$ and $z = 3$.

The galaxy spectral break induced by the cosmological Ly--$\alpha$
forest turns out to be the most effective feature to estimate
redshifts beyond $z \sim 2$. The so--called ``dropout'' technique has
been widely used to select high--redshift objects in current deep
surveys (Steidel \etal1995; Madau \etal1996). Notice though that for
faint objects the strong signature due to the IGM {\it external}
feature can be confused with the 4000 \AA\ Balmer break (which, on the
contrary is {\it intrinsic} to galaxy SED) resulting in a biased
redshift estimate.

\section{Matching the FSLY catalog}

The HDFN photometric catalog produced by FSLY will be used as a target
dataset for our analysis. It has been widely adopted in the recent
literature (see, for instance, Arnouts \etal1999; FSLY; Ben\'{\i}tez
2000) and consists of 1067 objects observed with HST WFPC2 in the four
photometric bands $U_{300}$, $B_{450}$, $V_{606}$, and $I_{814}$, and
with the IRIM camera at KPNO in the infrared range (Johnson $J, H$,
and $K$, Dickinson \etal(2000).  The FSLY catalog contains
fluxes, $f^\mathrm{obs}_i$, and errors, $\sigma_i$, for each galaxy in
all photometric bands ($i=1, \dots 7$). These quantities are those  
used in eqs.~(\ref{form_2})-(\ref{form_3}).

The availability of IR observations is of special relevance for our
analysis because, even for galaxies at $z \sim 4$, it allows us to
sample restframe emission at optical wavelengths, longward of the
Balmer break.  This greatly improves our chances to correctly
interpret galaxy colors and obtain a fair redshift estimate for the
most distant objects. As we can see in Fig.~\ref{u8}, within the
photometric errors, both elliptical and blue star-forming galaxies at
$z \sim 2$ show basically the same SED within the wavelength range
covered by the HST photometric system.  Only IR data can allow us to
break color degeneracy.

For our analysis we selected from the original FSLY dataset only those
objects with good photometry, that is with $S/N>5$ in at least two
bands, in order to rely confidently on at least one color.  Out of a
total of 1067 galaxies, we were left with 1041 objects in our
subsample.  As a result of our selection we will be biased against the
most distant objects beyond $z > 5$. For these galaxies the
{Ly-$\alpha$} break enters the $V_{606}$ band, making them detectable
only in the $I_{814}$ band (IR observations are not deep enough, in
any case, to grant a good $S/N$ detection for these distant objects).

The 4000 \AA\ Balmer break being one of the most useful spectral
feature for photometric redshift determination, we will single out a
``low--redshift'' subsample, including those galaxies for which the
break is still within the HST photometric bands (this happens for $z <
1.5$), and a ``high--redshift'' subsample ($z > 1.5$) for which the
4000 \AA\ break enters the IR photometric bands. 

\subsection{Internal uncertainty of reference libraries}

It is of special relevance for our aim to preliminarily assess the
effect of the observational uncertainties on the accuracy of redshift
determination with the SED fitting method.

We used a simple bootstrap procedure to generate ten
copies of the FSLY catalog, obtained by adding to each galaxy flux a
correction randomly extracted from the gaussian distribution of its 
photometric error.

For each of the reference template libraries we then compared, for
each object,. the redshift obtained from the measured photometry
$(z_{phot})$ and that from the bootstrapped galaxy photometry
$(z_{sim})$.

\begin{table}
\caption{Statistical uncertainty in redshift estimate due to the
photometric errors for each reference library (ten bootstrap
simulations).}
\label{tbl-sim}
\scriptsize
\begin{tabular}{lrcccc}

\hline
\large
 & &       &     &      &   \\
  & ${\bf \overline{\Delta z}}$ & ${\bf \sigma_z}$ & ${\bf \sigma_z}$ & ${\bf |\Delta z|>0.5}$ & ${\bf |\Delta z|>1.0}$ \\
 &  & ${\bf z < 1.5}$ & ${\bf z > 1.5}$    &   &   \\
 & &       &     &      &   \\
\hline
 & &       &     &      &   \\
 BUZ      & --0.01 & 0.16               & 0.24   &  11\% &  7\% \\
 & &       &     &      &  \\
 BC       & --0.01 & 0.16               & 0.23   &  12\% &  8\% \\
 & &       &     &      &   \\
 FRV      & --0.01 & 0.17               & 0.21   &  11\% &  7\% \\
 & &       &     &      &   \\
 CWWE     & 0.00   & 0.15               & 0.26   &  12\% &  8\%  \\
 & &       &     &      &   \\
\hline
\end{tabular}
\end{table}

\begin{figure*}
\resizebox{\hsize}{!}{\includegraphics{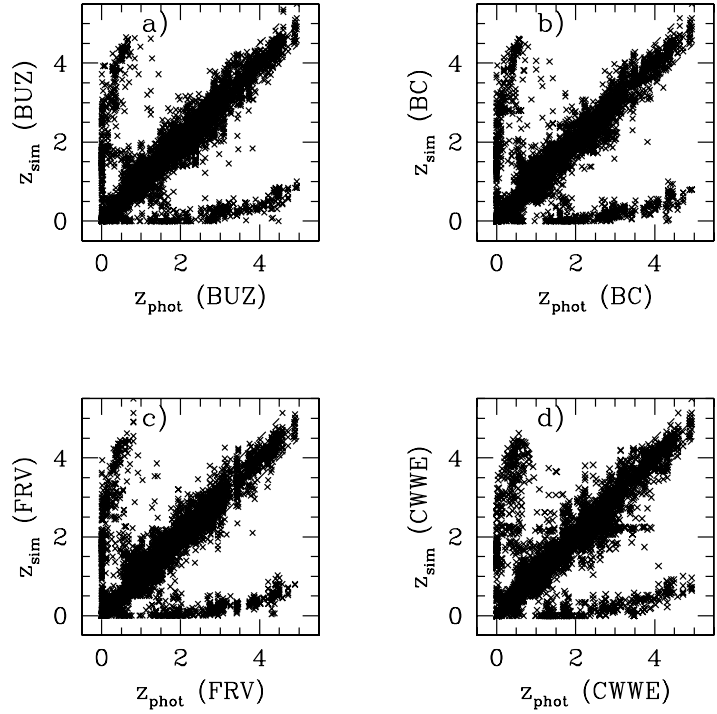}}
\caption{Comparison of the statistical uncertainty in the redshift
estimate due to photometric errors for the BC, BUZ and FRV reference
libraries, as well as for the CWWE empirical templates. Residuals are
from a bootstrap simulation of the FSLY galaxy catalog (1041 objects
for each of ten simulated catalog samples).}
\label{mock}
\end{figure*}

\noindent

We will eventually have:

\begin{equation}
\label{form_4bis}
\overline {\Delta z} ={\sum_{N_\mathrm{sim}}
{{(z_\mathrm{phot}-z_\mathrm{sim})}\over{N_\mathrm{sim}}}}\,,
\end{equation}

\noindent
and

\begin{equation}
\label{form_5bis}
\sigma_z^2 = \sum_{N_\mathrm{sim}} {{[z_\mathrm{phot}-z_\mathrm{sim}]^2}\over{N_\mathrm{sim}}}\,.
\end{equation}

\noindent
where $N_\mathrm{sim} = 10 \times 1041$ is the number of simulated
galaxies.

Moreover, we will check for the fraction of ``discrepant''
($|z_\mathrm{phot}-z_\mathrm{sim}|>0.5$) and ``catastrophic''
($|z_\mathrm{phot}-z_\mathrm{sim}|>1.0$) outliers in the low-- and
high--redshift galaxy subsamples.

Table~\ref{tbl-sim} gives a summary of the bootstrap results obtained
for each theoretical synthesis code as well as for the CWWE empirical
templates.  Figure~\ref{mock} shows in addition the distribution of
the redshift residuals.

Since photometric errors increase at fainter magnitudes, redshift
uncertainty is more pronounced for the most distant objects: $\sigma_z
\sim 0.16$ at $z<1.5$, while $\sigma_z \sim 0.25$ for the $z>1.5$
subsample. The catastrophic outliers are always less than 10\% of the
global sample, and there are no systematic effects, as expected.

In addition to the photometric errors, also physical degeneracies in
the $z$ vs. color space domain of the models could themselves result
in a source of numerical instability for the redshift estimates. In
this regard, different model sets could in principle provide also
different ``bias patterns'' in the redshift distribution.

Quite interestingly, Table~\ref{tbl-sim} shows that this is not the
case for the four template sets accounted here.  This might indicate
that all reference libraries contain the same kind of degeneracy. In
other words the galaxy colors are interpreted in the same way (from a
statistical point of view) by the different template sets.  

\begin{figure*}
\resizebox{\hsize}{!}{\includegraphics{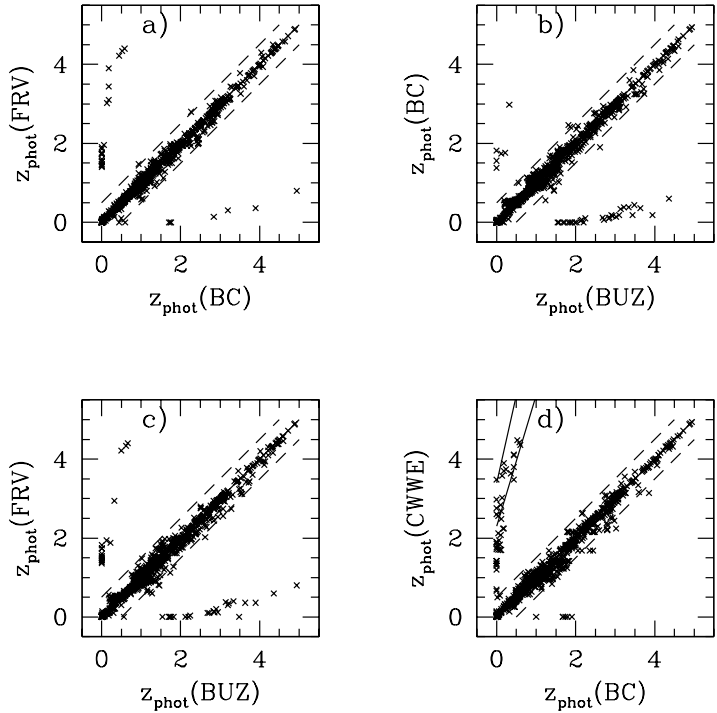}}
\caption{Comparison of HDFN expected redshift distribution, for
different reference libraries. The solid line is for $\Delta z=0.0$,
while long--dashed strip is for $\Delta z = \pm 0.5$. The vertical
strip (solid lines) in panel {\it d} encloses the catastrophic
outliers coming from the misinterpretation of the Balmer/Lyman break
(see text for further details).}
\label{fig-2}
\end{figure*}

In Figure~\ref{mock} the distribution of redshift residuals for
catastrofic outliers ($|\Delta z|>1.0$) has a well defined pattern,
and it is actually possible to identify two main sources for this
pattern. The first is the color degeneracy between redshift $z \sim 0$
and $ 1.3 < z < 2.1$, that, given the photometric bands available in
FSLY, shows up in all the templates. In the redshift interval $ 1.3 <
z < 2.1$, the entire Lyman series has not yet entered the $U_{300}$
band (cf. Fig.~\ref{fig-1}), while the Balmer break is located between
the $I_{814}$ and $J$ bands. The uncertainty in the photometric
observations can easily cause a featureless objects to move within the
ranges $z \sim 0$ and $ 1.3 < z < 2.1$.  The second major source of
the pattern visible in Figure~\ref{mock} are the biased $z$ estimates
occuring when, due to large photometric errors, the Balmer break is
mistaken for the Lyman decrement or viceversa.  When the photometric
errors in the IR bands are large enough to diluite/enhance a
real/fictitious Balmer decrement, the objects affected are shuffled
between the redshift ranges $ 0 < z < 1$ and $z > 2.5$.

\subsection{Comparison among reference libraries}

We can now compare photometric redshifts obtained using different
reference template libraries.

Given two independent redshift estimates, namely $z_\mathrm{phot, 1}$
and $z_\mathrm{phot, 2}$ from different libraries, we can again define
a simple statistical moment of the residual distribution in the form:

\begin{equation}
\label{form_4}
\overline {\Delta z} ={\sum_{N_\mathrm{g}} {{(z_\mathrm{phot,
1}-z_\mathrm{phot, 2})}\over{N_\mathrm{g}}}}\,,
\end{equation}

\noindent
while a variance can be computed as

\begin{equation}
\label{form_5}
\sigma_z^2 = \sum_{N_\mathrm{g}} {{[z_\mathrm{phot, 1}-z_\mathrm{phot,
2}]^2}\over{N_\mathrm{g}}}\,,
\end{equation}

\noindent
where $N_\mathrm{g} = 1041$ is the number of galaxies.

Also in this case, we should check for the fraction of discrepant
($|z_\mathrm{phot, 1}-z_\mathrm{phot, 2}|>0.5$) and catastrophic
($|z_\mathrm{phot, 1}-z_\mathrm{phot, 2}|>1.0$) outliers in the low--
and high--redshift galaxy subsamples.

In addition, we should also test whether $\overline {\Delta z}$ is
consistent with zero or there is some systematic drift in data
residuals.  Mean standard deviation should be explored against
possible trends as a function of galaxy redshifts.

Figure~\ref{fig-2} and Table~\ref{tbl-1} summarize the relevant
results. The agreement between BC, BUZ and FRV (panel {\it a--c} in
Fig.~\ref{fig-2}) seems quite good, with less than 5\% of catastrophic
outliers, and $\sigma_z$ below 0.15 over the entire redshift range.

\begin{table}
\caption{Comparison of redshift distributions from the
different reference libraries} \label{tbl-1}

\scriptsize
\begin{tabular}{lrcccc}

\hline
   &   &  &       &     &         \\
   & ${\bf \overline{\Delta z}}$ & ${\bf \sigma_z}$ & ${\bf \sigma_z}$ & ${\bf |\Delta z|>0.5}$ & ${\bf |\Delta z|>1.0}$\\
   &   & ${\bf z < 1.5}$  & ${\bf z > 1.5}$  &     &         \\
\hline
   &   &  &       &     &         \\
 BC vs. FRV               & 0.02    & 0.09               & 0.12  & 3\% &  3\% \\
   &   &  &       &     &         \\
 BC vs. CWWE              & 0.03     & 0.12               & 0.17  & 6\%  & 5\% \\ 
   &   &  &       &     &         \\
 BUZ vs. BC               & 0.00    & 0.12               & 0.13  & 4\% &  3\% \\
   &   &  &       &     &        \\
 BUZ vs. FRV              & 0.02    & 0.13               & 0.15  & 5\%  & 4\% \\
   &   &  &       &     &         \\
 BUZ vs. CWWE              & 0.02    & 0.15               & 0.19  & 6\%  & 4\% \\
   &   &  &       &     &         \\
 FRV vs. CWWE              & 0.01    & 0.12               & 0.18  & 6\%  & 5\% \\ 
   &   &  &       &     &         \\
\hline
\end{tabular}
\end{table}

The comparison with the CWWE template set gives slightly poorer values
for $\sigma_z$ (cf. panel {\it d} in Fig.~\ref{fig-2}), especially for
the most distant galaxies in the FSLY sample, but again a small
fraction of discrepant redshift objects.

As we can appreciate in Table~\ref{tbl-1}, there are only marginal
systematic drifts in the $\overline {\Delta z}$ residuals.

It should be noted that in the synthetic libraries the ``primeval''
evolutionary scenario is described with the same templates (the four
starburst models from Leitherer \etal1999).  On the other hand, even
by replacing, for example, the ``primeval'' BC library with the young
galaxies provided by the BC code itself, results of the comparison
with other models are left almost unchanged.

Comparing the results of Table ~\ref{tbl-1} with those of Table
~\ref{tbl-sim}, it is clear that the whole SED fitting procedure is
largely insensitive to any specific choice of the model reference
library. Rather than depending on the detailed galaxy SED, the
redshift is mainly constrained by the recognition of the Balmer and/or
the Ly--$\alpha$ break in the galaxy flux distribution.

\begin{figure}
\resizebox{\hsize}{!}{\includegraphics{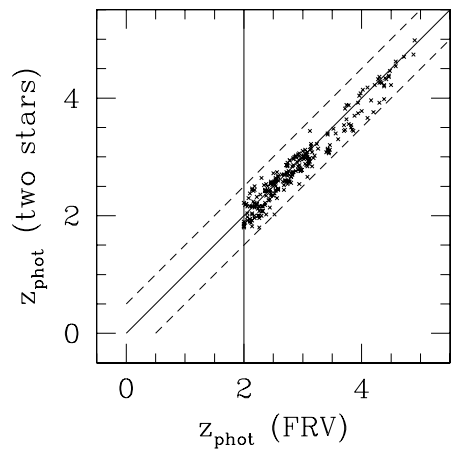}}
\caption{SED fitting of the FSLY galaxy subsample beyond $z>2$. The
data distribution by matching the FRV template set is compared with a
``minimal'' reference library consisting of two Kurucz' (1992) model
atmospheres for stars of spectral type $B0$ and $B2$.  The relative
$\sigma_z$ results 0.18 with no apparent drift in the data
distribution.  The solid line is for $\Delta z=0.0$, while
long--dashed strip is for $\Delta z = \pm 0.5$. See text for a full
discussion.}
\label{stelle}
\end{figure}

As a quite extreme but instructive example, in Fig.~\ref{stelle} we
tried a fit of the galaxy subsample with fiducial redshift $ z > 2 $
(259 objects, as estimated in {\it every one} of our reference
libraries) using as a template set two Kurucz' (1992) model
atmospheres for stars of $T = 27\,000$ and $18\,000\ \mathrm{K}$.
These temperatures are roughly consistent with a $B0$--$B2$ spectral
type like for stars of 10--20 $M_\odot$ that dominate starburst galaxy
luminosity in the early 10--100 Myr of life (Leitherer \etal1999). The
redshift distribution relying on this ``minimal'' template set well
agrees with that obtained using, for example, the FRV model library
($\sigma_z =0.18$ and no catastrophic outliers). At high redshifts,
therefore, the SED fitting method provides robust estimates of galaxy
redshifts, even when using an extremely poor library of templates.  It
is clear, however, that such a simplified procedure cannot be of any
utility if we are interested in a more accurate analysis of
high--redshift data, such as to investigate age distribution or galaxy
morphology at different cosmological times. 

Another interesting check of internal consistency of the SED fitting
method deals with the age--redshift constraint. We discarded such
constraint, by allowing the procedure to choose also galaxy models
older than the Hubble time. Plotting the galaxy age information and
the redshift information, both obtained from the photometric redshift
algorithm, will allow to check for their reciprocal consistency.

\begin{figure}
\resizebox{\hsize}{!}{\includegraphics{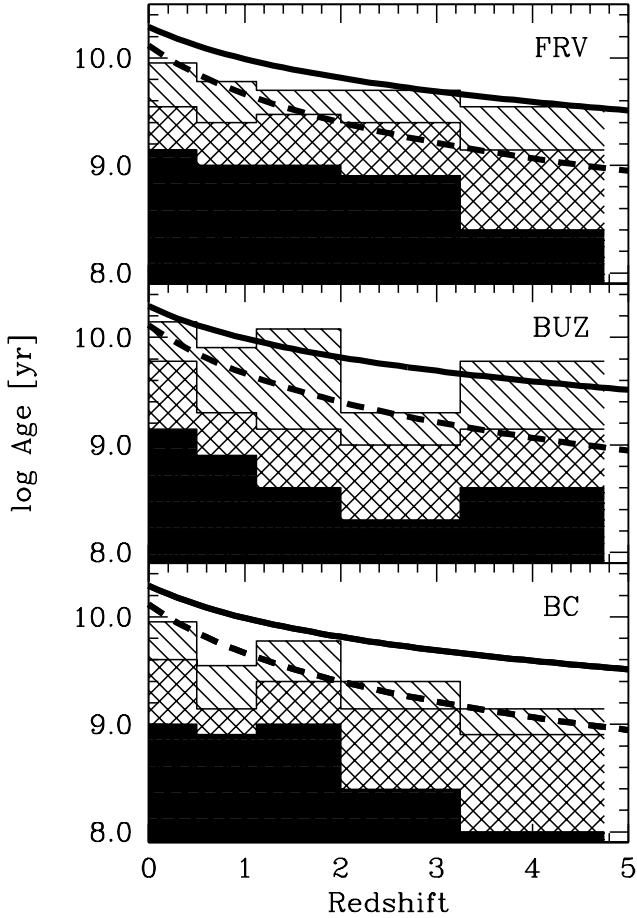}}
\caption{The expected age--redshift relation from the BC, BUZ and FRV
best templates by imposing no cosmological constraint. HDFN galaxies
have been grouped in order to guarantee at least 100 objects per
redshift bin.  Solid, grid, and diagonal-shaded histograms are the
50\%, 75\%, and 90\% envelopes of galaxy age distribution.  In each
plot, solid and dotted lines are the theoretical age-$z$ relation for
$H_0 = 50$ and $q_0 = 0.0$ and 0.5, respectively.}
\label{fig-6}
\end{figure}

For this exercise we grouped HDFN galaxies
such as to guarantee at least 100 objects per redshift
bin. Fig.~\ref{fig-6} reports our results according to the BC, BUZ and
FRV codes.

Over the whole sample, the percentages of the fitting templates
older than the actual Hubble time ($[H_0, q_0] = [50, 0.0]$) with BC,
BUZ and FRV codes are 3.9 \%, 8.5 \% and 5.7 \% respectively.  For
each theoretical code the empirical redshift-age relation is virtually
compatible with any cosmological model with $H_0 = 50$ and $q_0 <
0.5$.

\subsection{Catastrophic outliers}

Although wide general agreement exists among template libraries in
predicting galaxy redshift, it is worth investigating in some detail
the few striking cases of catastrophic discrepancies (i.e. $|\Delta
z|>1.0$).

There are basically two groups of such outliers, as in the 
distribution of redshift residuals due to photometric errors. The
first group consists of objects with featureless spectrum, that is
with no significant break signature (within the photometric
errors). This is typically the case of the faintest blue objects in
the HDFN sample.  Due to color degeneracy between redshift $z \sim 0$
and $ 1.3 < z < 2.1$, these galaxies are shuffled within the two
intervals by different libraries.  The redshift interval $ 1.3 < z <
2.1$ is intrinsically critical for the SED fitting algorithm given the
photometric bands available to us.  As already noted above, while the
entire Lyman series has not yet entered the $U_{300}$ band
(cf. Fig.~\ref{fig-1}), the Balmer break is located between the
$I_{814}$ and $J$ bands. Therefore, redshift estimate entirely relies
on the recognition of the smooth SED, and little changes in the
reference templates can easily lead to diverging data interpretations.

The second class of catastrophic outliers comprises those objects for
which the Balmer break is mistaken for the Lyman decrement or
viceversa, according to the template library chosen, generating a
series of biased $z$ estimates beyond $z > 2.5$. This effect is
especially evident when comparing CWWE vs. BC (cf. panel {\it d} in
Fig.~\ref{fig-2}). A too scanty sampling of galaxy morphology, like in
the CWWE templates, is the cause of the relative exacerbation of the
effect. For example galaxies with no deep Balmer break and relatively
blue optical colors can be interpreted as Lyman break galaxies, in
absence of alternative intermediate age elliptical templates.

\begin{figure}
\resizebox{\hsize}{!}{\includegraphics{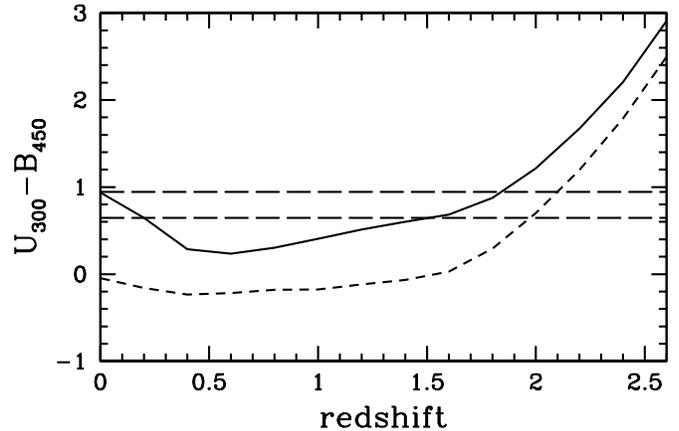}}
\caption{The $U_{300}-B_{450}$ apparent color as a function of
redshift for the CWW $Im$ empirical template (solid line) and
Leitherer et al. (1999) 50 Myr starburst model (dashed line). In
absence of any starburst template, a fraction of star--forming
galaxies at $z \sim 2$ (with $U_{300}-B_{450} \sim 0.8$) might be
interpreted as local ($z \leq 0.2$) irregulars.}
\label{col}
\end{figure}

More in general, a major source of catastrophic errors in the SED
fitting procedure can result from the lack of starburst templates in
the galaxy reference library. Standard Hubble types alone (even
including $Im$ irregulars) are not ``blue'' enough to reproduce the
colors of the most active star--forming objects seen at high
redshift. In Fig.~\ref{col} we show the $U_{300} - B_{450}$ apparent
color of the CWWE $Im$ template (solid line) and that of the Leitherer
et al. (1999) 50 Myr starburst model (dotted line) as a function of
redshift. Lacking starburst templates can introduce a systematic shift
in the redshift estimate ($\Delta z \sim 0.3$) for those galaxies with
$(U_{300}-B_{450})\sim 0.8$. Even worst, in some cases one might
erroneously confuse UV--enhanced galaxies at $z \sim 2$ with local
irregulars sampled in their optical SED. When comparing FSLY
photometric redshifts with spectroscopic redshifts both this effects
are evident. 

\section{Starbursts and the role of ISM and IGM}

Dust reddening and IGM absorption have a large impact on the detection
of starburst--galaxies at high redshifts, as these effects can
effectively masque the original starburst spectrum.

In a simple experiment using the CWWE library and imposing $E(B-V) =
0$ in our fit, about 20\% of galaxies in the FSLY catalog can suitably
be matched by (reddening--free) starburst templates. When dust
attenuation is accounted for, and $E(B-V)$ is left to vary as a free
fitting parameter, this fraction raises to about $55\%$ of the whole
galaxy sample. A confident evaluation of the fraction of UV--enhanced
galaxies at the different distances would have a pervasive impact in a
number of cosmological problems, like for instance the study of the
cosmic SFR (Madau \etal1996; Steidel \etal1999; Thompson \etal2000).

\begin{figure}
\resizebox{\hsize}{!}{\includegraphics{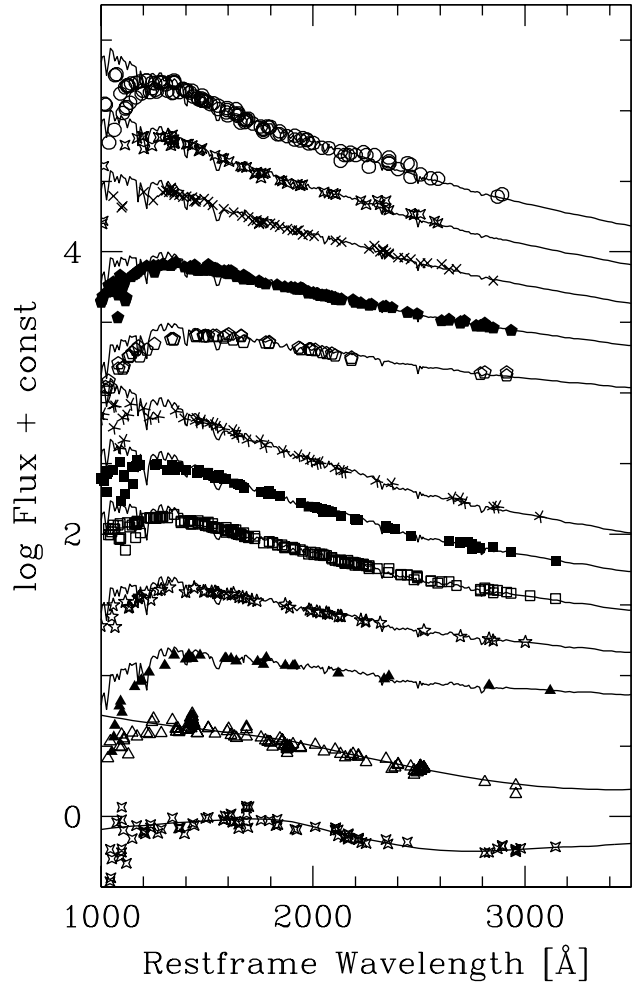}}
\caption{Restframe composite SED from the HST data for the 341 HDFN
galaxies with fiducial $z>1.5$ compared with the CWWE reference
templates. From top to bottom solid lines display the starburst
template with $t=500$ Myr and $E(B-V)=0.0,0.05,0.1,0.2,0.3$, the $t =
50$ Myr starburst model with $E(B-V)=0.0,0.05,0.1,0.2,0.3$, the $Scd$
and the $Im$ CWW spectra.}
\label{fig-7}
\end{figure}

To explore the selective influence of interstellar dust on the SED
fitting procedure, we considered the FSLY subsample with fiducial $z >
1.5$ (as estimated in {\it every one} of our reference libraries) and
compared it with the CWWE empirical library (see Fig.~\ref{fig-7}).

Out of 348 selected galaxies, 341 can be fitted by starburst
templates, $Scd$ spirals or later morphological types.

In a dust--free fitting scheme (i.e. by assuming $E(B-V) = 0$) we
derive the following relative partition among fiducial morphological
types:

\begin{center}
[others : Scd : Im : Starbursts] = [7 : 34 : 160 : 147]
\end{center}

\noindent
with about $12\%$ objects poorly fitted, that is at least 3--$\sigma$
off in each band according to eq.~(\ref{form_2}). Incidentally, if
we exclude from the template library the starburst models we obtain:

\begin{center}
[others : Scd : Im] = [7 : 43 : 298]
\end{center}

\noindent
but at the cost of almost doubling the fraction of poorly fitted
objects ($\sim 24 \%$ of the total sample).

If we leave $E(B-V)$ as a free fitting parameter (in the range $0.0
\leq E(B-V) \leq 0.3$), the quality of the fit improves remarkably
(only 5\% of objects deviates by 3--$\sigma$ or more), and an even
larger number of starburst candidates is obtained:

\begin{center}
[others : Scd : Im : Starbursts] = [7 : 25 : 41 : 275].
\end{center}

\begin{figure}
\resizebox{\hsize}{!}{\includegraphics{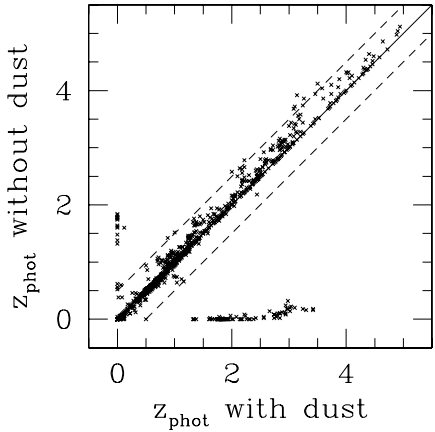}}
\caption{The selective influence of galaxy internal reddening on the
SED fitting performance. The HDFN redshift distribution from the BC
template set is displayed with and without taking into account for ISM
absorption. This is done by leaving $E(B-V) = 0$ or as a free fitting
parameter in the range $0.0 \leq E(B-V) \leq 0.3$. Long--dashed lines
in the plot represent $\Delta z= \pm 0.5$.}
\label{dust_nodust}
\end{figure}

\noindent
A dust--free fitting scheme also drastically increases the number of
catastrophic outliers ($\sim 10 \%$ on the entire catalog,
cf. Fig.~\ref{dust_nodust}), and at $z>1.5$ the redshift of reddened
galaxies is overestimated by $\overline {\Delta z} \sim 0.2$ since IGM
absorption, which is an increasing function of redshift, is forced to
mimic color reddening.

In Fig.~\ref{fig-7} the scatter of the HST data (blueshifted according
the $z_\mathrm{phot}$ value and normalized according the scaling
factor $s$ of eq.~\ref{form_3}) around the CWWE templates is shown
directly. As we can see, at $\lambda > 1216$ \AA\ the best templates
are an accurate description of high--redshift galaxy
photometry. Notice however that at $\lambda < 1216$ \AA\ the observed
fluxes deviate from the best templates: the experimental data do
contain the IGM absorption, while the best templates do not.

\begin{figure}
\resizebox{\hsize}{!}{\includegraphics{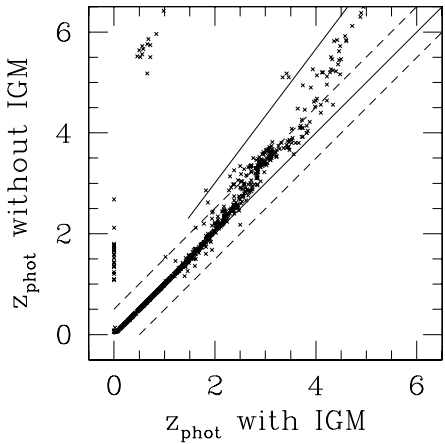}}
\caption{Comparison of the HDFN redshift distribution obtained from
the BUZ reference library with and without taking into account the IGM
absorption. Long--dashed lines in the plot represent $\Delta z= \pm
0.5$. The solid line shows the expected upper limit in the
redshift drift due to the the misinterpretation of the intrinsic Lyman
break with the Ly--$\alpha$ forest effect.}
\label{igm_noigm}
\end{figure}

The effect of IGM at high redshifts is explored in
Fig.~\ref{igm_noigm}. Redshifts in the FSLY catalogue are determined
using the BUZ reference library with and without IGM taken into
account.  A drift clearly appears beyond $z > 1.5$ when the
Ly--$\alpha$ break enters the $U_{300}$ band. The shift is generated
by the wavelength difference between the 912 \AA\ Lyman break in the
galaxy spectrum and the various steps of the Ly--$\alpha$ forest.  The
fraction of outliers amounts to about $12 \%$ on the entire catalog
and to $\sim 21 \%$ of the $z>1.5$ subsample.

\section{Discussion and conclusions}

We investigated the stability of photometric redshift estimates in the
SED fitting method with respect to the galaxy template set chosen.  We
used as a target dataset the FSLY catalog in the HDFN and as libraries
of galaxy colors those produced by the theoretical codes of Bruzual \&
Charlot (1993), Buzzoni (2000) and Fioc and Rocca--Volmerange (1997),
as well as the empirical template set of Coleman \etal (1980). This
way our model reference framework comprises (and enlarges) the sample
of template libraries used by almost all the literature on the field.

The existence of non negligible photometric errors in the data is by
itself an important source of uncertainty in the redshift
estimates. Comparing the FSLY results with a number of bootstrap
simulation of the same galaxy sample we estimate that the internal
uncertainty in the SED fitting performance amounts to $\sigma_z \sim
0.16$ at $z<1.5$ and $\sigma_z \sim 0.25$ at $z>1.5$, with $\sim 8 \%$
of catastrophic outliers. This result is almost independent from the
adopted model library and can be read as a lower limit to the
intrinsic accuracy of the fitting method given the photometric errors
of the FSLY catalogue.

The external redshift uncertainty, found by comparing results obtained
using different theoretical models, is typically $\sigma_z \sim 0.11$
at $z<1.5$ and $\sigma_z \sim 0.13$ at $z>1.5$, with less than $4 \%$
of catastrophic outliers. Comparing theoretical models and the CWWE
empirical set we obtained $\sigma_z \sim 0.13$ at $z<1.5$ and
$\sigma_z \sim 0.18$ at $z>1.5$, with less than $5 \%$ of catastrophic
outliers. Our results show that, given the photometric errors in the
data, the SED fitting procedure is to a large extent almost
insensitive to any specific choice of the model reference library.

The lack of starburst templates in the model library, on the other
hand, introduces a systematic bias in the redshift estimate ($\Delta z
\sim 0.3$) and a sensible increase of catastrophic outliers as
UV--enhanced galaxies at $z \sim 2$ can be confused with local
irregulars sampled in their optical SED.

Not including in the fitting procedure ISM dust reddening or IGM
opacity has a dramatic impact on the distribution of redshifts
measured.  

A dust--free fitting scheme drastically increases the number of
catastrophic outliers ($\sim 10 \%$ on the entire catalog) and at
$z>1.5$ the redshift of dust reddened galaxies is overestimated by
$\overline {\Delta z} \sim 0.2$.  When dust attenuation is accounted
for, the fraction of galaxies matched by starburst templates raises
from $\sim 20\%$ (by imposing $E(B-V) = 0$) to $\sim 55\%$ of the
whole HDFN sample.

In an IGM--transparent fitting scheme a drift in the redshift
estimates clearly appears beyond $z > 1.5$ and the fraction of
outliers amounts to $\sim 21 \%$ of the high--redshift subsample.

Our results suggest that, especially at high redshifts, the
description of physical processes of photon absorption in the
interstellar and in the intergalactic medium plays a key role in the
SED fitting method for interpreting photometric data. Any specific
choice of the templates, as long as these describe both normal and
starburst galaxies, has on the contrary a minor effect on the
performance of the method.

\begin{acknowledgements}
It is a pleasure to thank S. Charlot for useful
suggestions, and the referee, R.I. Thompson, for his 
comments. M. Massarotti acknowledges financial support by
Osservatorio Astronomico di Brera and Fondazione Cariplo.
\end{acknowledgements}

\end{document}